\documentclass[aps,onecolumn,noshowpacs,nobalancelastpage,tightenlines,floatfix]{revtex4}
\usepackage{amssymb}
\usepackage{graphicx}
\usepackage[dvips]{epsfig}
\usepackage{dcolumn}
\usepackage{bm}
\usepackage{amsmath}

\begin{document}

\title{Observation of Strong Coupling between One Atom and a Monolithic
Microresonator}
\author{Takao Aoki$^{a}$, Barak Dayan, E. Wilcut, W. P. Bowen$^{b}$, A. S.
Parkins$^{c}$, and H. J. Kimble}
\affiliation{Norman Bridge Laboratory of Physics 12-33, California Institute of
Technology, Pasadena, California 91125, USA}
\author{T. J. Kippenberg$^{d}$ and K. J. Vahala}
\affiliation{T. J. Watson Laboratory of Applied Physics, California Institute of
Technology, Pasadena, California 91125, USA}
\date{\today}
\maketitle

\textbf{Over the past decade, strong interactions of light and matter at the
single-photon level have enabled a wide set of scientific advances in
quantum optics and quantum information science. This work has been performed
principally within the setting of cavity quantum electrodynamics \cite%
{miller05,berman94,walther04,haroche05} with diverse physical systems \cite%
{vahala-review}, including single atoms in Fabry-Perot resonators \cite%
{miller05,nussmann05}, quantum dots coupled to micropillars and photonic
bandgap cavities \cite{khitrova06,badolato05}, and Cooper-pairs interacting
with superconducting resonators \cite{wallraff04,chiorescu04}. Experiments
with single, localized atoms have been at the forefront of these advances
\cite{mckeever03b,mckeever04,keller04,birnbaum05,legero04} with the use of
optical resonators in high-finesse Fabry-Perot configurations \cite{rempe92}%
. As a result of the extreme technical challenges involved in further
improving the multilayer dielectric mirror coatings \cite{hood01} of these
resonators and in scaling to large numbers of devices, there has been
increased interest in the development of alternative microcavity systems 
\cite{vahala-review}. Here we show strong
coupling between individual Cesium atoms and the fields of a high-quality
toroidal microresonator. From observations of transit events for single
atoms falling through the resonator's evanescent field, we determine the
coherent coupling rate for interactions near the surface of the resonator. 
We develop a theoretical model to quantify
our observations, demonstrating that strong coupling is achieved, with the
rate of coherent coupling exceeding the dissipative rates of the atom and
the cavity. Our work opens the way for investigations of optical processes
with single atoms and photons in lithographically fabricated
microresonators. Applications include the implementation of quantum networks
\cite{cirac97,briegel00}, scalable quantum logic with photons \cite{duan04},
and quantum information processing on atom chips \cite{treutlein06}.}\newline

The realization of large-scale quantum networks \cite{cirac97,briegel00}
requires the capability to inter-connect many `quantum nodes', each of which
could consist of a microresonator containing a set of trapped atoms. The `%
quantum channels' to connect these nodes would be optical fibres, with
strong interactions in cavity quantum electrodynamics (QED) providing an
efficient interface between light and matter. Here we provide a critical
step towards a feasible quantum network by demonstrating strong coupling of
single atoms to microresonators fabricated on Silicon wafers in large
numbers by standard lithographic techniques followed by a laser-reflow
process \cite{armani03}. Combined with the capability to couple light
efficiently to and from such cavities directly via a tapered optical fibre
\cite{spillane03}, toroidal microcavities offer promising capabilities for
new nonlinear interactions of single atoms and photons across distributed
networks.

Our efforts follow the pioneering work of V. Braginsky \textit{et al.} \cite%
{braginsky89} and later studies \cite{vernooy98} by employing the
whispering-gallery modes of fused silica microtoroidal resonators \cite%
{spillane05}. As depicted in Fig. 1, a Silicon chip containing a collection
of $35$ microtoroidal resonators is located inside a vacuum chamber at $%
10^{-9}$ Torr and is positioned to couple a particular resonator to a
tapered fibre. The toroids have major diameter $D\simeq 44$ $\mu $m and
minor diameter $d\simeq 6$ $\mu $m \cite{spillane05}. By judicious choice of
the point of contact between the surface of the resonator and the tapered
fibre, we attain critical coupling, in which the forward propagating power $%
P_{F}$ in the fibre drops to near zero for the probe frequency $\omega _{p}$
equal to the cavity resonance frequency $\omega _{C}$ \cite{spillane03}.
Measurements of the cavity transmission in the absence of atoms are
presented in Fig. 2. Note that the forward flux $P_{F}$ and associated
transmission spectrum $T_{F}$\ are analogous to the reflected flux and
reflection spectrum from a Fabry-Perot cavity \cite{spillane03}.
By varying the temperature of the Silicon chip, the detuning $\Delta
_{AC}\equiv \omega _{C}-\omega _{A}$ between $\omega _{C}$ and the atomic
resonance at $\omega _{A}$ ($6S_{1/2},F=4\longrightarrow 6P_{3/2},F^{\prime
}=5^{\prime }$ transition in Cesium) can be controlled with uncertainty $%
\simeq \pm 2$ MHz (see Appendix A).

Cold atoms are delivered to the vicinity of the toroidal resonator from a
small cloud of Cesium atoms cooled to $T\simeq 10$ $\mu $K and located $10$
mm above the Silicon chip. Every $5$ seconds, the cloud is dropped,
resulting in about $2\times 10^{6}$ atoms in a $3$ mm ball at the height of
the chip, with then a few dozen atoms passing through the external
evanescent field of the toroidal resonator. By way of two single-photon
detectors $(D_{F1},D_{F2})$ (see Appendix A), we continuously monitor the
forward propagating signal $P_{F}$ from a frequency-stabilized probe beam $%
P_{in}$ coupled to the toroidal resonator. The interaction of each
individual atom with the evanescent field destroys the condition of critical
coupling, leading to an increase in $P_{F}$. The measurement cycle then
repeats itself for $2.5$ seconds for a reference measurement, this time with
no atomic cloud formed above the microtoroid.

\begin{figure}[tb]
\includegraphics[width=10cm]{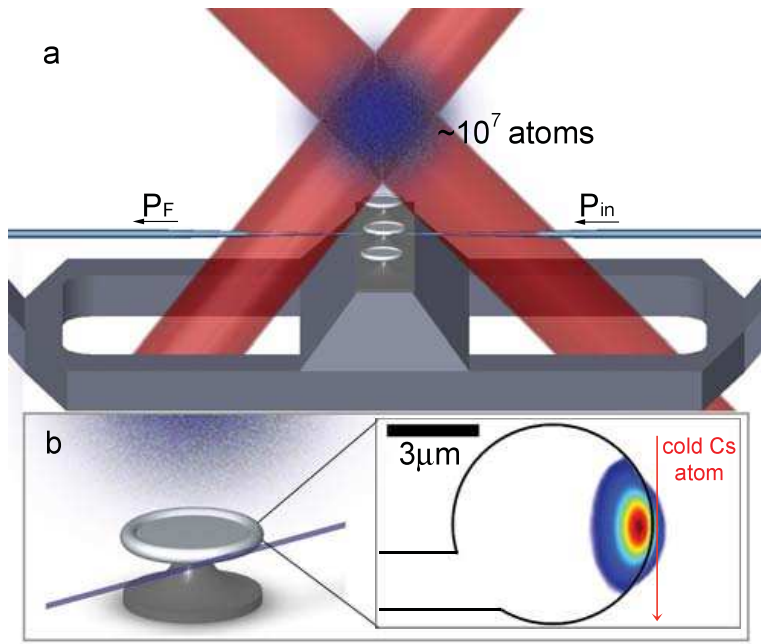}
\caption{\textbf{Simple diagram of the experiment. (a)} A cloud of cold
Cesium atoms and associated trapping lasers above an array of microtoroidal
resonators. Light from the probe beam $P_{in}$ is coupled into a resonator
by way of the fibre taper, with the forward propagating output $P_{F}$
coupled into the taper. \textbf{(b)} Illustration of an SiO$_{2}$
microtoroidal resonator, fibre taper, and atom cloud above. The calculated
field distribution for the lowest order resonator mode is shown by the color
contour plot on the right. Cold Cesium atoms fall through the external
evanescent field of this mode and are thereby strongly coupled to the
resonator's field.}
\label{fig1}
\end{figure}

Figure 3 displays typical records $C(t)$ for the number of single-photon
detection events within time bins of $\delta t=2$ $\mu $s as functions of time
$t $ for the forward signal $P_{F}(t)$. Measurements are displayed with
(Fig. 3(a)) and without (Fig. 3(b)) atoms for the case of equal probe and
cavity frequencies, $\omega _{p}=\omega _{C}$, for $\Delta _{AC}\approx 0$,
and with mean intracavity photon number $\bar{n}_{0}\approx 0.3$ for the
forward circulating mode of the toroidal resonator $a$ (see Appendix A). The
traces in both Fig. 3(a) and Fig. 3(b) exhibit background levels that result
from the nonzero ratio $P_{F}/P_{in}\sim 0.005$ at critical coupling in the
absence of atoms. However, Fig 3(a) clearly evidences sharp peaks of
duration $\Delta t\approx 2$ $\mu $s for the forward propagating light $P_{F}(t)$, 
with an individual peak shown more clearly in the inset. Each event
arises from the transit of a single atom through the resonant mode of the
microtoroid, with about $30$ events per cycle observed. Fig. 3(c) examines
the temporal profile of transit events in more detail by way of the cross
correlation $\Gamma (\tau )$ of photoelectric counts $%
C_{1}(t_{1}),C_{2}(t_{1}+\tau )$ from the detectors $(D_{F1},D_{F2})$ for $%
P_{F}$ (see Appendix A). This result agrees reasonably well with the
theoretical prediction for atom transits through the calculated field
distribution shown in Fig. 1(b).

By applying a threshold requiring $C(t)\geq 6$ counts for $C(t)$ as in Fig.
3(a, b), we find the average time dependence $\bar{C}_{\geq 6}(t)$ over
about 100 measurement cycles. Fig. 3(d) displays the results both with and
without atoms, with the average counts $\Sigma _{6}(t)$ derived from $\bar{C}%
_{\geq 6}(t)$ by summing over successive time bins $\delta t=2$ $\mu $s for $%
1$ ms intervals. The peak in transit events is consistent with the expected
distribution of arrival times for atoms dropped from our atom cloud. By
contrast, negligible excess events (i.e., $C(t)\geq 6$) are recorded for the
cases without atoms.

Focusing attention to the central region indicated by the dashed lines in
Fig. 3(d), we examine in Fig. 3(e) the probability $P(C)$ to record $C$
counts within $\delta t=2$ $\mu $s. Evidently, when the atom cloud is
present, there is a statistically significant increase (of at least fifteen
sigma) in the number of events with $C\geq 4$. These are precisely the
events illustrated by the inset in Fig. 3(a) and the cross correlation in
Fig. 3(c), and are associated with single atom transits near the surface of
the toroidal resonator. By varying the value of $\bar{n}_{0}$, we have
confirmed that the large transit events evident in Fig. 3 are markedly
decreased for $\bar{n}_{0}\gtrsim 1$ photon, which indicates the saturation
of the atom-cavity system.

\begin{figure}[tb]
\begin{center}
\includegraphics[width=10cm]{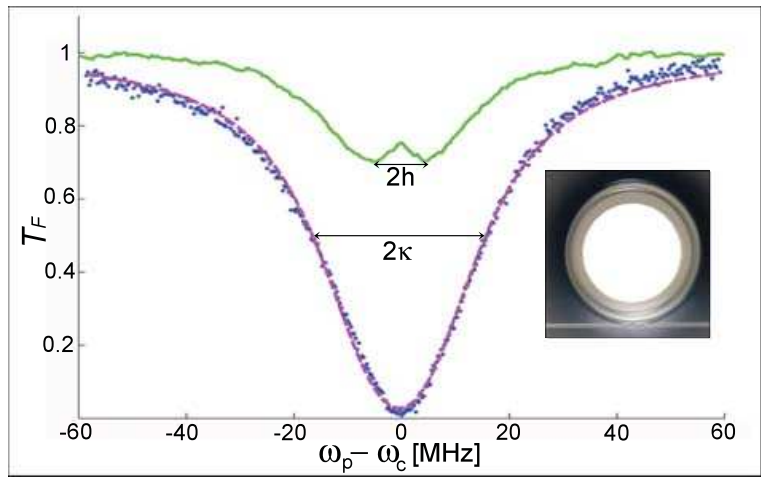}
\end{center}
\caption{\textbf{Cavity transmission function $T_{F}=P_{F}/P_{in}$ as a
function of probe frequency $\protect\omega _{p}$}. The lower trace is taken
for critical coupling, and the upper trace for conditions of under coupling
\protect\cite{spillane03}. From fits to such traces for critical coupling
(red dashed curve), we find $(\protect\kappa ,h)/2\protect\pi =(17.9\pm
2.8,4.9\pm 1.3) $ MHz, with $\protect\kappa$, $h$ being the overall cavity
field decay rate and the scattering-induced coupling between two
counter-propagating modes of the microtoroid, respectively (see Appendix for
more details). Inset -- Photograph of a microtoroid and coupling fibre.}
\label{new-fig2}
\end{figure}

A quantitative description of our observations in Fig. 3 of individual atom
transits requires the development of a new theoretical model in cavity QED.
In the appendix, we present such a model and show that the underlying
description of the interaction of an atom with the fields of the toroidal
resonator is in terms of normal modes $(A,B)$ (see Fig. 5 in Appendix B),
which have mode functions $\psi _{A,B}(\rho ,x,z)$ that are standing waves $%
(\cos kx,\sin kx)$ around the circumference $x$ of the toroid, with $\rho $
the radial distance from the surface and $z$ the vertical coordinate. $\psi
_{A,B}(\rho ,x,z)$ have a calculated peak coherent coupling $g_{0}/2\pi$ of
70 MHz for the lowest order modes of our resonator (such as that illustrated
in Fig. 1(b)). The normal modes A,B result from the coupling of two
oppositely directed travelling waves by scattering at rate $h$, with the
resulting mode splitting manifest in Fig. 2. Note that the presence of two
normal modes leads to a $\sqrt{2}$ increase in the coupling constant in our
case as compared to the one predicted by the Jaynes-Cummings model for an
atom interacting with a single traveling-wave mode (see Appendix B for
further details).

Guided by this theory, we have performed a series of measurements similar to
those presented in Fig. 3 to determine the coherent coupling rate $g_{0}$
for interactions of single atoms with our toroidal resonator, but now with
various values of the atom-cavity detuning $\Delta _{AC}$, keeping the probe
resonant with the cavity: $\omega _{p}\approx \omega _{C}=\omega
_{A}+\Delta _{AC}$. The qualitative idea is that large single-atom
transit events will occur only over a range of detunings $\Delta _{AC}$
determined by $g_{0}$. Specifically, the decrease in the forward
transmission $T_{F}$ induced by atom transits as a function of $\Delta _{AC}$
is described by a Lorentzian with width $\beta $ set by $g_{0}$ (see
Appendix B). In our case, $g_{0}=g_{0}(\rho ,x,z)\approx g_{0}(\rho ,x,Vt)$,
where $V$ is the velocity of the dropped atoms in the $z$ direction. Thus, a
numerical integration was performed over $\rho $, $x$, and $t$ to derive the
theoretical expectation for $T_{F}(\Delta _{AC})$, presented in Fig. 4(a)
for three values of $g_{0}^{m}$, where $g_{0}^{m}$ is the maximal coupling
that an atom can experience in its interaction with the cavity modes.
Indeed, we see that the width $\beta $ grows monotonically with $g_{0}^{m}$.
However, the average value of $T_{F}$ is not a parameter that is readily
measured in our current experiment, in which we expect many short individual
transits, some of which are too weak to be distinguished from the background
noise (Fig. 3(e)). A parameter that describes our actual experimental
measurements more closely is the probability to obtain a transit which
results in transmission above a certain threshold. The two measures are
closely related, such that this probability decreases with detuning $\Delta
_{AC}$ in the same fashion as does $T_{F}$.

Figure 4(b--d) presents the results of our measurements for the average
number of transit events per atom drop, $N_{drop}^{av}(C\geq C_{0})$, which
have photoelectric counts greater than or equal to a threshold value $C_{0}$
for a set of seven detunings $\Delta _{AC}$. In accord with the expectation
set by Fig. 4(a), there is a decrease in the occurrence of large transit
events for increasing $\Delta _{AC}$ in correspondence to the decrease in
the effective atom-cavity coupling coefficient for large atom-cavity
detunings. The full curves shown in Fig. 4(b--d) are the results of
theoretical calculation for these measurements, with the relevant cavity
parameters $(\kappa ,h)$ determined from fits as in Fig. 2.

We first ask whether the data might be explained by an effective value $%
g_{0}^{e}$ for the coherent coupling of atom and cavity field, without
taking into account the fact that individual atoms transit at radial
distances $\rho $ which vary from atom to atom. Fig. 4(b) examines this
possibility for various values of $g_{0}^{e}$, assuming a coupling
coefficient $g_{0}^{e}\psi _{A,B}(x)=g_{0}^{e}[\cos kx,\sin kx]$ averaged
along one period in $x$ (as in Fig. 4(a)). Apparently, an effective value $%
g_{0}^{e}/2\pi \approx 40$ MHz provides reasonable correspondence between
theory and experiment for large events $C\geq 6$.


\begin{figure}[tb]
\begin{center}
\includegraphics[width=10cm]{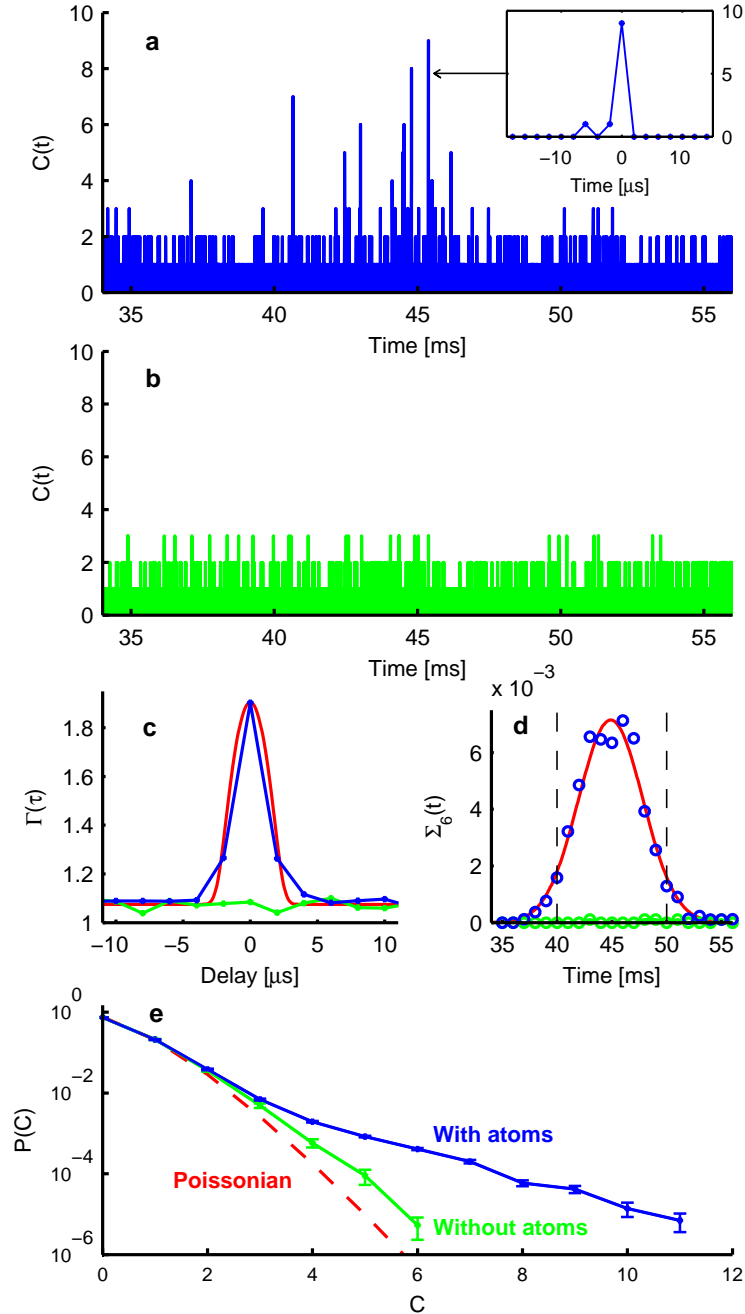}
\end{center}
\caption{\textbf{Measurements of the forward signal $P_F$ in the presence of
falling atoms (blue) and without atoms (green). (a, b)} Single-photon
counting events $C(t)$ as a function of time $t$ after the release of the
cold atom cloud at $t=0$, with \textbf{(a)} and without \textbf{(b)} atoms
dropped. $C(t)$ gives the total number of counts recorded for time bins of $%
\protect\delta t=2$ $\protect\mu $s duration. The inset in \textbf{(a)}
shows the time profile for a single-atom transit. \textbf{(c)} Normalized
cross-correlation $\Gamma (\protect\tau )$ of the forward signal counts from
two detectors $(D_{F1},D_{F2})$ showing the time profile associated with
atom transit events. The smooth (red) curve is the theoretically predicted
average cross correlation for a transit event with one atom, taking into
account drop height of 10 mm and the spatial shape of the mode, as depicted
in Fig. 1(b). \textbf{(d)} Counts $\Sigma _{6}(t)$ obtained from $\bar{C}%
_{\geq 6}(t)$ by summing over $1$ ms intervals, compared to a Gaussian
distribution which fits the rate of atom transits assuming a 3 mm (FWHM)
cloud of cold atoms dropped from 10 mm above the microtoroid. \textbf{(e)}
Probability $P(C)$ to detect $C$ counts within $\protect\delta t=2$ $\protect%
\mu $s bins for the central interval shown by the vertical dashed lines in
\textbf{(d)}, compared with Poissonian statistics (red) with the same mean
number of counts ($\sim$ 0.25 per 2 $\protect\mu$s). The excess probability
above the poissonian level in the no atoms case is predominately due to
instability in the cavity temperature, which results in small fluctuations
in the forward flux. Error bars show $\pm$1 s.d.}
\label{fig3}
\end{figure}

We adapt our theory to the actual situation of atoms arriving randomly at
radial and circumferential coordinates by introducing a mesh grid over $%
(\rho ,x)$, and then computing the cavity transmission function $T_{F}(t)$\
from $\psi _{A,B}(\rho ,x,z(t))$ for atomic trajectories $z(t)$ over this
grid. We account for the time resolution $\delta t=2$ $\mu $s of our data
acquisition by a suitable average of $T_{F}(t)$\ over such time bins (as was
also true in Fig. 4(b)). The results from these calculations are displayed
in Fig. 4(c--d) as the set of full curves for three values of coherent
coupling $g_{0}$ for the cavity mode functions $\psi _{A,B}(\rho ,x,z)$,
where in Fig. 4(b--d) the theory is scaled to match the measured $%
N_{drop}^{av}(C\geq C_{0})$ at $\Delta _{AC}=0$. From such comparisons, we
determine the maximal accessible $g_{0}^{m}/2\pi =(50\pm 12)$ MHz.
Note that this conclusion is insensitive to the choice of cutoff $C_{0}$
over the range $4\leq C_{0}\leq 9$ for which we have significant transit
events. Strong coupling with $g_{0}^{m}>(\kappa ,\gamma )$ is thereby
achieved, where $(\kappa ,\gamma )/2\pi $ = (17.9 $\pm $ 2.8, 2.6) MHz are
the dissipative rates for the cavity field and the atom.

According to our calculations, $g_{0}^{m}/2\pi$ = 50 MHz corresponds to the
coupling rate expected at a distance of roughly 45 nm from the surface of
the microtoroid. We estimate that due to the attractive van der Waals forces
(see Ref~\cite{courtois96}), atoms which enter the evanescent field with a
distance $\rho\leq45 $ nm from the microtoroid are expected to strike its
surface in less than 1 $\mu$s, thus preventing such atoms from generating
appreciable transit events in the transmission function $T_F$.

\begin{figure}[tb]
\begin{center}
\includegraphics[width=10cm]{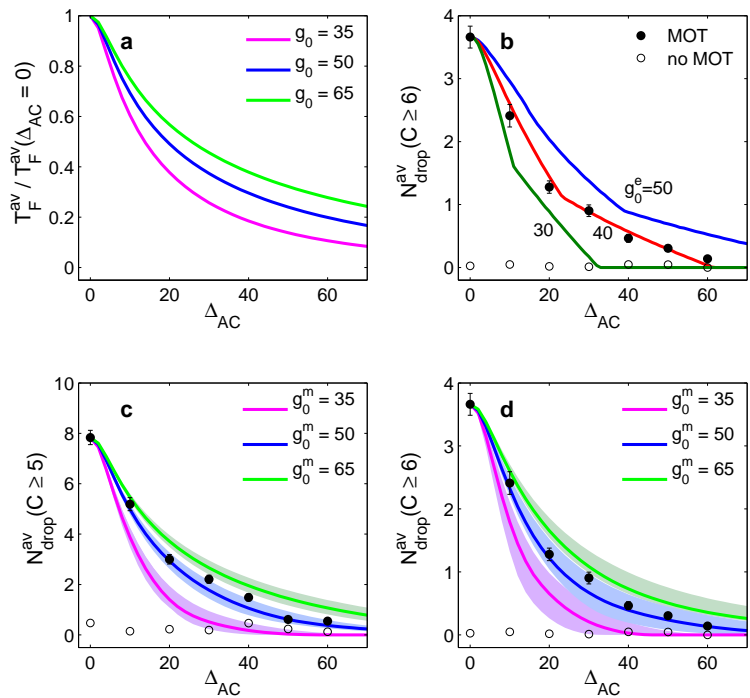}
\end{center}
\caption{\textbf{Measurements of transit events as a function of the
atom-cavity detuning $\Delta _{AC}$}. Events are shown in the presence
of atoms (filled circles) and without atoms (empty circles), compared with
the theoretical calculations (lines). \textbf{(a)} Theoretical calculation
for the average of the transmission $T_{F}(\protect\omega _{p}=\protect%
\omega _{C})$ as a function of $(\Delta _{AC},g_{0})$. Red, $g_{0}$=35;
blue, $g_{0}$=50; green, $g_{0}$=65. \textbf{(b-d)} Measurements for the
average number of events per drop of the atom cloud $N_{drop}^{av}(C\geq
C_{0})$ plotted against the atom-cavity detuning $\Delta _{AC}$, with $%
C_{0}=6$; \textbf{(b, d)} and $C_{0}=5$ \textbf{(c)}. Error bars show $\pm $%
1 s.d. The data are taken for probe frequency $\protect\omega %
_{p}\approx \protect\omega _{C}=\protect\omega _{A}+\Delta _{AC}$. The
full curves are theoretical results as discussed in the text. The widths of
the curves are determined from the experimental uncertainties in $(\protect%
\kappa ,h)$. \textbf{(b)} Theory for $N_{drop}^{av}(C\geq 6)$ without radial
averaging to deduce an effective coupling $g_{0}^{e}/2\protect\pi =40$ MHz.
Theory for \textbf{(c)} $N_{drop}^{av}(C\geq 5)$ and \textbf{(d)} $%
N_{drop}^{av}(C\geq 6)$ with radial and azimuthal averaging leading to $%
g_{0}^{m}/2\protect\pi =50$ MHz. Red, $g_{0}^{m}$=35; blue, $g_{0}^{m}$=50;
green, $g_{0}^{m}$=65.}
\label{fig4}
\end{figure}

In summary, we report the first observation of strong coupling for
single atoms interacting with an optical resonator other than a conventional
Fabry-Perot cavity. The monolithic microtoroidal resonators \cite%
{armani03} employed here have the capability of input-output coupling
with small parasitic losses, with a demonstrated ideality of more than $%
99.97\%$ \cite{spillane03}. Moreover, quality factors $Q=4\times 10^{8}$
have been realized at $\lambda =1550$ nm \cite{kippenberg04}\ and $Q\simeq
10^{8}$ at $\lambda =850$ nm \cite{spillane05}, with good prospects for
improvements to $Q\simeq 10^{10}$ \cite{vernooy98a}. For these parameters,
the efficiency for coupling single photons into and out of the resonator
could approach $\epsilon \sim 0.99-0.999$ \cite{spillane03}, while still
remaining firmly in the regime of strong coupling \cite{spillane05}. Such
high efficiency is critical for the realization of scalable quantum networks
\cite{cirac97,briegel00} and photonic quantum computation \cite{duan04}.
Indeed, of the diverse possibilities for the distribution and processing of
quantum information with optical cavities \cite%
{vahala-review,khitrova06,badolato05}, the system of single atoms coupled to
microtoroidal resonators arguably provides one of the most promising
avenues. Beyond efficient input-output coupling \cite{spillane03}, strong
coupling to a material system with long-lived internal states has now
been achieved, although here in a primitive, proof-of-principle
setting. An outstanding technical challenge is to trap single atoms near the
surface of the microtoroid, with one possibility having been investigated in
ref. \cite{vernooy97}.

\textbf{Acknowledgements --} We thank M. Eichenfield, K. W. Goh and S. M.
Spillane for their contributions to the early stages of this experiment, and
T. Carmon, A. Gross and S. Walavalker for their contributions to the current
realization. The work of H.J.K. is supported by the National Science
Foundation, the Disruptive Technology Office of the Department of National
Intelligence, the Army Research Office, and Caltech. The work of
K.J.V. is supported by DARPA, the Caltech Lee Center and the National
Science Foundation. B.D., W.P.B. and T.J.K. acknowledge support as Fellows
of the Center for the Physics of Information at Caltech. E.W. acknowledges
support as a Ford Predoctoral Fellow from the US National Academies. A.S.P.
acknowledges support from the Marsden Fund of the Royal Society of New
Zealand.

\appendix

\section{Experimental Details}

\textit{Preparation and characterization of cold atoms} -- Each measurement
cycle in our experiment takes about $2.5$ sec, and includes approximately $2$
seconds for loading a magneto-optical trap (MOT), followed by $20$ ms of
polarization-gradient cooling of the atoms (with the magnetic fields for the
MOT turned off). The trapping and cooling beams are then switched off and
the atoms fall on the microtoroid.

For each run, we measure the number and arrival times of atoms in the
falling atom cloud $2$ mm above the microtoroid with a laser beam resonant
with the $6S_{1/2},F=4\longrightarrow 6P_{3/2},F^{\prime }=4^{\prime }$
transition \cite{yu01}. In each cycle we observe approximately 30 atom
transits during the center 10 ms of our data-collection time window. This
value is in a reasonable agreement with the theoretically calculated rate of
20 transits, which was derived by comparing the measured density of the
falling atom cloud to the numerically calculated interaction area of the
evanescent field. Every cycle with cold atoms is followed by an identical
cycle with no trapped atoms for which the magnetic field for the MOT is
turned off during the loading period. We also carried out other tests for
the \textquotedblleft no atoms\textquotedblright\ case, including switching
off the repumping light for the MOT (with all other parameters unchanged).

\textit{Excitation and detection system} -- The frequency $\omega _{\mathrm{p%
}}$ of the probe beam $P_{\mathrm{in}}$ in Fig. 1(a) is actively stabilized
to a fixed detuning from the atomic resonance $\Delta _{\mathrm{A}}=\omega _{%
\mathrm{A}}-\omega _{\mathrm{p}}$ via saturation spectroscopy to within $\pm
100$ kHz. The cavity resonance at $\omega _{\mathrm{C}}$ is monitored
relative to $\omega _{\mathrm{A}}$ and $\omega_{\mathrm{p}}$ for each drop
of the atom cloud and each reference cycle, and is controlled by
temperature-tuning a thermoelectric device upon which the Silicon chip is
mounted, with a frequency shift of approximately 3 MHz per mK of temperature
change. Data is automatically recorded whenever the condition of critical
coupling (i.e. $P_{\mathrm{F}}<1\%$ of the maximal value of 30 counts per 2 $%
\mu$s) was achieved, corresponding to $\omega _{\mathrm{p}}=\omega _{\mathrm{%
C}}$ to within $\pm 2$ MHz. Note that in our experiment we use blue detuning
of the cavity and the probe relative to the atom, since red detuning could
lead to a resonant interaction with the 6S$_{1/2}$ F=4 $\rightarrow$ 6P$%
_{3/2}$ F$^{\prime}$ = 4$^{\prime}$ transition at large atom-cavity detuning
$\Delta _{\mathrm{AC}}\equiv \omega _{\mathrm{C}}-\omega _{\mathrm{A}}$.
Additionally, blue detuning leads to an under-estimation of $g$, as the
dipole forces become increasingly repulsive as we blue-detune our cavity,
possibly leading to a more rapid reduction in the probability of transits.
However, our calculations indicate that in our current experimental settings
the light forces are significantly smaller than the van der Waals forces
over the entire relevant interaction region of the atom with the evanescent
field, and their effect on the atom's motion and temperature is small. This
situation is mostly due to the small populations of the dressed states that
are coupled to the atom for excitation with $\omega _{\mathrm{p}} \approx
\omega _{\mathrm{C}}$. Specifically, in our experiment the probe field
frequency is always tuned to the empty cavity resonance, between the two
vacuum-Rabi sidebands which correspond to the two dressed states of the atom
and the coupled cavity mode (as described in Fig. 5). Thus, as the energy
splitting between these states (and hence the associated dipole potential)
grows during the atom transit, their population drops dramatically (and so
does their contribution to the mechanical potential experienced by the
atom), as they decouple from the probe field, leading to the described
increase in the forward flux $T_{\mathrm{F}}$. The resulting forces are of
the order of few MHz/$\mu$m, leading to estimated displacements of up to 40
nm during the 2 $\mu$s of atom transit for intermediate values of $\Delta _{%
\mathrm{AC}}$. Further measurements to explore the effect of light forces
along with its inclusion in the theoretical model are in progress.

The probe $P_{\mathrm{in}}$ enters the vacuum apparatus by way of a
single-mode fibre through a Teflon feedthrough \cite{abraham98}. This fibre
is spliced to the fibre taper. The forward propagating signal $P_{\mathrm{F}}
$ in Fig. 1(a) exits the vacuum chamber in the same fashion. $P_{\mathrm{F}}$
is then directed to a 50:50 fibre beam splitter whose outputs are detected
by a pair of single-photon counting modules (SPCMs) $(D_{\mathrm{F1}},D_{%
\mathrm{F2}})$ each with overall quantum efficiency $\alpha \simeq 0.5$ and
dark counts $<100$ per second. Using two detectors enables us to avoid, or
at least assess, phenomena that are related to the non-ideality of these
detectors, such as their $\sim $50 ns dead time, saturation at fluxes
exceeding a few million counts per second, \textquotedblleft after-pulsing"
effects which may result in a false second count, etc. The two
detectors allow the detection of photons that are separated temporally by
less than the dead time of these detectors (which are not number-resolving),
and increase the maximal flux of photons that can be detected by our
system before saturation effects take place. This method also enables
further analysis of the photon statistics of the light, such as the
cross-correlation between the two series of photon counts presented in Fig.
3(c).

Detection events from $(D_{\mathrm{F1}},D_{\mathrm{F2}})$ are time-stamped
relative to the drop time of the atom cloud, and stored for later analysis.
The data in the figures refers to the total counts from the combined outputs
of $(D_{\mathrm{F1}},D_{\mathrm{F2}})$. The overall propagation efficiency $%
\xi $ from the fibre taper at the position of the toroidal resonator to the
input beam splitter for $(D_{\mathrm{F1}},D_{\mathrm{F2}})$ is $\xi =0.70\pm
0.02$. An additional SPCM was used to monitor the backwards flux, namely
light that was coupled into the cavity, scattered into the
counter-propagating mode and then transmitted backwards into the taper. When
atom transits occurred, the observed increase in the forward flux was
accompanied by a decrease in this backward flux.

In the absence of an atom, the average intracavity photon number is $\bar{n}%
_{\mathrm{0}}\simeq 0.3$ for the forward propagating mode ($a$) for critical
coupling at $\omega_{\mathrm{p}}=\omega _{\mathrm{C}}$. If the probe is then
detuned such that $|\omega _{\mathrm{p}}-\omega _{\mathrm{C}}|\gg \kappa $,
the average number of counts recorded in a $2$ $\mu $s interval is $C_{%
\mathrm{\Delta \gg \kappa }}\approx 30$, which provides a calibration of the
flux $P_{\mathrm{F}}$ given the known propagation and detection losses.

\section{Theory for a two-level atom coupled to two toroidal modes}

To understand our observations in quantitative terms, we have developed a
theoretical model for a two-level atom interacting with the quantized fields
of the toroidal resonator. The two-level atom has transition frequency $%
\omega _{\mathrm{A}}$ and raising and lowering operators $\sigma ^{\pm }$.
The two counter-propagating modes of the toroidal resonator are taken to be
degenerate with common frequency $\omega _{\mathrm{C}}$ (in the absence of
scattering), and are described by annihilation (creation) operators $a$ ($%
a^{\dag }$) and $b$ ($b^{\dag }$), respectively. In our actual resonators,
modes are coupled due to scattering with a strength that is parameterized by
$h$. A coherent probe of frequency $\omega _{\mathrm{p}}$ in the input field
of the fibre taper, $a_{\mathrm{in}}$, couples to mode $a$ with a strength $%
\mathcal{E}_{\mathrm{p}}$. The input field to mode $b$ is taken to be vacuum
as in our experiments. In a frame rotating at the probe frequency $\omega _{%
\mathrm{p}}$, a simple Hamiltonian that models our system is thus \cite%
{rosenblit04}
\begin{eqnarray}
H/\hbar  &=&\Delta _{\mathrm{A}}\sigma ^{+}\sigma ^{-}+\Delta \left( a^{\dag
}a+b^{\dag }b\right) +h\left( a^{\dag }b+b^{\dag }a\right)   \notag \\
&&+\left( g_{\mathrm{tw}}^{\ast }a^{\dag }\sigma ^{-}+g_{\mathrm{tw}}\sigma
^{+}a\right) +\left( g_{\mathrm{tw}}b^{\dag }\sigma ^{-}+g_{\mathrm{tw}%
}^{\ast }\sigma ^{+}b\right)   \notag \\
&&+\left( \mathcal{E}_{\mathrm{p}}^{\ast }a+\mathcal{E}_{\mathrm{p}}a^{\dag
}\right) ,
\end{eqnarray}%
where $\Delta _{\mathrm{A}}=\omega _{\mathrm{A}}-\omega _{\mathrm{p}}$, $%
\Delta =\omega _{\mathrm{C}}-\omega _{\mathrm{p}}$. The coherent interaction
of the atom with the evanescent traveling-wave fields of the $a,b$ modes is
described by $g_{\mathrm{tw}}=g_{\mathrm{0}}^{\mathrm{tw}}\,\phi _{\mathrm{tw%
}}^{\pm }(\rho ,x,z)$, where the mode functions $\,\phi _{\mathrm{tw}}^{\pm
}(\rho ,x,z)=$ $f(\rho ,z)e^{\pm ikx}$ with $f(\rho ,z)\sim e^{-\alpha \rho }
$ ($\alpha \sim 1/\lambdabar $). The coordinates
$(\rho ,x,z)$ are derived from cylindrical coordinates $(r,\theta ,z)$,
where $r$ is the radial distance from the axis of symmetry of the toroid,
with then $\rho =r-D/2$; $\theta $ is the azimuthal angle in a plane
perpendicular to the symmetry axis, with $x=r\theta $ as the position around
the circumference of the toroid; and $z$ is the vertical dimension along the
symmetry axis. $k$ is the vacuum wave vector. The field decay rate for the
resonator modes is $\kappa =\kappa _{\mathrm{i}}+\kappa _{\mathrm{ex}}$,
where $\kappa _{\mathrm{i}}$ represents intrinsic losses and $\kappa _{%
\mathrm{ex}}$ describes extrinsic loss due to (adjustable) coupling of the
modes to the fibre taper \cite{spillane03,gorodetsky00,kippenberg02}. The
atomic excited state population decays with rate $2\gamma $.

\begin{figure}[tb]
\includegraphics[width=10cm]{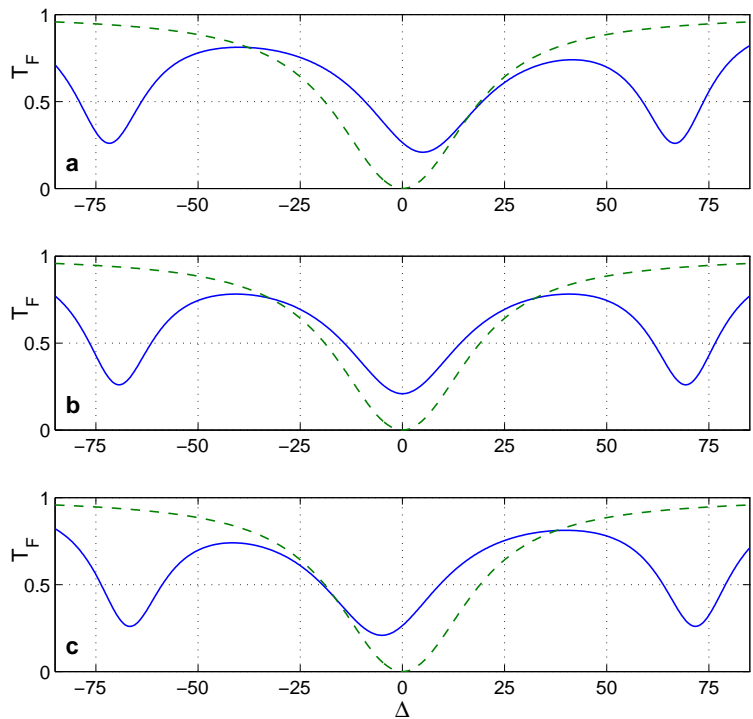}
\caption{Calculated spectra $T_{\mathrm{F}}$ of the forward flux in the
presence of an atom (solid blue line) as a function of probe detuning $%
\Delta $ for different $x$-coordinates around the circumference of the
toroid, namely (a) $kx=0$, (b) $kx=\protect\pi /4$, (c) $kx=\protect\pi /2$.
In all cases, $\Delta _{\mathrm{AC}}=0$ and $g_{\mathrm{0}}/2\protect\pi =70$
MHz as appropriate for an atom at the external surface of our toroid. Also
shown as the dashed curve is the spectrum in the absence of an atom ($g_{%
\mathrm{0}}=0$) under the same conditions of critical coupling. $T_{\mathrm{F%
}}(\protect\omega _{\mathrm{p}})$ is normalized by the probe flux far off
resonance, with $(\protect\kappa ,h)$ determined from fits as in Fig. 2.}
\label{fig5}
\end{figure}

The output field in the forward direction is given by $a_{\mathrm{out}}=a_{%
\mathrm{in}}+\sqrt{2\kappa _{\mathrm{ex}}}\,a$ \cite{gardiner85}. Assuming
only weak excitation of the atom, we can compute the output photon flux from
this relationship and a linearized approximation to the equations of motion
for the atomic coherence $\langle \sigma ^{-}\rangle $ and field amplitudes $%
\langle a\rangle $ and $\langle b\rangle $. Characteristic spectra $%
T_{F}(\omega _{\mathrm{p}})$ of the forward flux $P_{\mathrm{F}}$ as a
function of probe detuning and for different $x$-coordinates of the atom are
shown in Fig.~\ref{fig5}, where $|g_{\mathrm{tw}}|/2\pi =70/\sqrt{2}$ MHz
and $\Delta _{\mathrm{AC}}=0$. The spectra are normalized by the flux of a
far-off-resonant probe field. Also shown is the spectrum in the absence of
an atom ($g_{\mathrm{tw}}=0$), under the (experimental) conditions of
critical coupling; specifically, where $\kappa _{\mathrm{ex}}=\kappa _{%
\mathrm{ex}}^{\mathrm{cr}}=\sqrt{\kappa _{\mathrm{i}}^{2}+h^{2}}$, for which
$P_{\mathrm{F}}(\Delta =0)\approx0$ \cite%
{spillane03,gorodetsky00,kippenberg02}.

The coupled-atom spectra shown in Fig.~\ref{fig5} can be understood by
considering the normal modes of the microtoroidal resonator, $A=(a+b)/\sqrt{2%
}$ and $B=(a-b)/\sqrt{2}$, in terms of which the Hamiltonian can be written
\begin{eqnarray}
H &=&\Delta _{\mathrm{A}}\sigma ^{+}\sigma ^{-}+(\Delta +h)A^{\dag
}A+(\Delta -h)B^{\dag }B  \notag \\
&&+\frac{1}{\sqrt{2}}\left[ \mathcal{E}_{\mathrm{p}}^{\ast }\left(
A+B\right) +\mathcal{E}_{\mathrm{p}}\left( A^{\dag }+B^{\dag }\right) \right]
\notag \\
&&+g_{A}\left( A^{\dag }\sigma ^{-}+\sigma ^{+}A\right)  \notag \\
&&-ig_{B}\left( B^{\dag }\sigma ^{-}-\sigma ^{+}B\right) .
\label{eq:H_normalmodes}
\end{eqnarray}

The coherent coupling for the $(A,B)$ modes is given here by $g_{\mathrm{A,B}%
}=g_{\mathrm{0}}\,\psi _{\mathrm{A,B}}(\rho ,x,z)$, where $g_{\mathrm{0}}=%
\sqrt{2}g_{\mathrm{0}}^{\mathrm{tw}}$. The mode functions $\psi _{\mathrm{A,B%
}}(\rho ,x,z)$ for the normal modes of the cavity are $\psi _{\mathrm{A}%
}(\rho ,x,z)=$ $f(\rho ,z)\cos (kx)$ and $\psi _{\mathrm{B}}(\rho ,x,z)=$ $%
f(\rho ,z)\sin (kx)$. Significantly, the underlying description of the
interaction of an atom with the toroidal resonator is thus in terms of
standing waves $\psi _{\mathrm{A,B}}(\rho ,x,z)$ along the surface of the
toroid. The splitting for the normal modes $(A,B)$ induced by scattering $h$%
\ is displayed for undercoupling to our resonator in Fig. 2.

With reference to Fig.~\ref{fig5}(a,c), we see that for $kx=0~(\pi /2)$ the
atom couples only to mode $A$ ($B$) of frequency $\omega _{\mathrm{C}}+h$ ($%
\omega _{\mathrm{C}}-h$) with strength $g_{\mathrm{0}}$, leading to a
pronounced reduction in $T_{\mathrm{F}}$ at probe detunings $\Delta \simeq
-h\pm g_{\mathrm{0}}$ ($h\pm g_{\mathrm{0}}$) for the case $\Delta _{\mathrm{%
AC}}=0$ shown, i.e., at the \textquotedblleft vacuum-Rabi
sidebands\textquotedblright . The central feature in the transmission
spectrum $T_{\mathrm{F}}$\ at $\Delta =h$ ($-h$) is the spectrum of the
uncoupled normal mode $B$ ($A$). By contrast, at $kx=\pi /4$ in (b), the
atom couples with equal strength to both normal modes and a system of three
coupled oscillators is realized (in the linear approximation), the normal
mode frequencies of which occur at $\omega _{\mathrm{C}}$ and $\sim (\omega
_{\mathrm{C}}\pm g_{\mathrm{0}})$ for $\Delta _{\mathrm{AC}}=0$.\newline

\begin{figure}[tb]
\includegraphics[width=13cm]{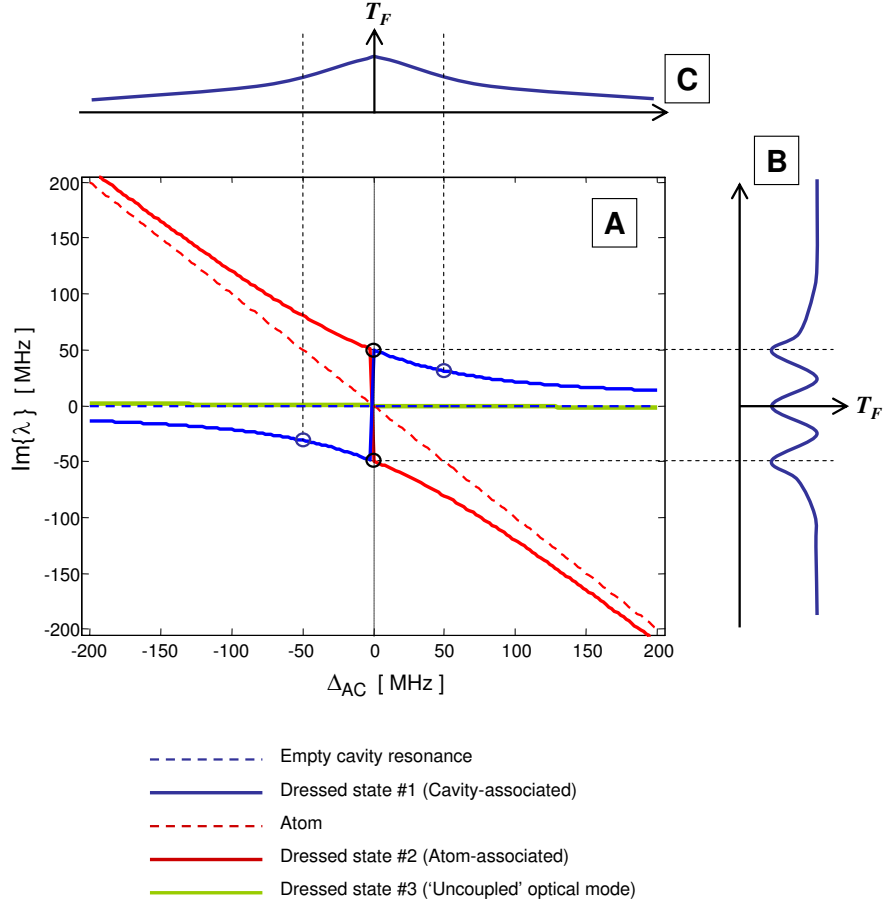}
\caption{(A) The three calculated eigenvalues of atom-toroid coupled system
for ($g$,$\protect\kappa$,$h$)=(50,18,5) MHz. (B,C) Qualitative
illustrations of the transmission $T_{\mathrm{F}}$, (B) at fixed atom-cavity
detuning $\Delta_{\mathrm{AC}}=0$ as a function of the probe frequency $%
\protect\omega_{\mathrm{p}}$, and (C) as a function of $\Delta_{\mathrm{AC}}$%
, with probe frequency fixed to be $\protect\omega_{\mathrm{p}} = \protect%
\omega_{\mathrm{C}}$. }
\label{fig6}
\end{figure}

The eigenvalues of the atom-toroid coupled system are drawn in Fig. 6(A)
relative to the cavity resonance (blue, dashed), using the same parameters
as Fig 5(b). In this representation the atom (red, dashed) has positive
energy detuning for negative values of $\Delta_{\mathrm{AC}}$, and negative
detuning for positive values of $\Delta_{\mathrm{AC}}$. The coupling with
the atom lifts the degeneracy between the atom and the cavity at $\Delta_{%
\mathrm{AC}}$, leading to the expected splitting of $2g$ between dressed
states \#1 and \#2. However, since the atom is never fully coupled to both
normal modes (A,B), a third dressed state (\#3, green) remains practically
unchanged by the atom. This three mode structure is represented in both Fig.
5(a-c) and Fig. 6(B), which illustrate the transmission $T_{\mathrm{F}}$ at
fixed $\Delta_{\mathrm{AC}}=0$ as a function of the probe frequency,
demonstrating minima at 0 and $\pm g$. Similarly, the coupling strength
manifests itself in the dependence of $T_{\mathrm{F}}$ on $\Delta_{\mathrm{AC%
}}$, as shown in Fig 6(C) for probe frequency fixed to the cavity resonance,
as is the case in our experiment. Thus, the decrease in the forward flux $T_{%
\mathrm{F}}$ at the cavity resonance $\omega _{\mathrm{p}}=\omega _{\mathrm{C%
}}$ as a function of the cavity-atom detuning $\Delta_{\mathrm{AC}}$ is a
generic feature of the eigenvalue structure of the system.

\begin{figure}[h]
\begin{center}
\includegraphics[scale=0.6]{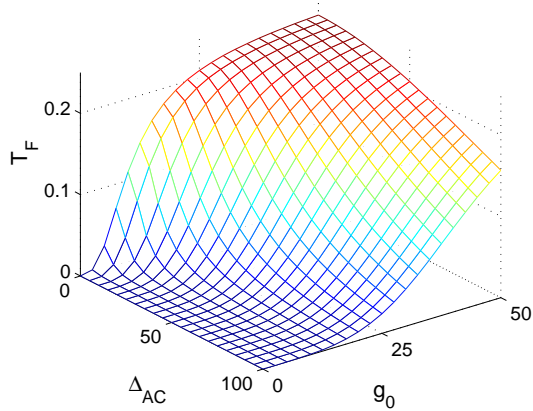}
\end{center}
\caption{Theoretical calculation for transmission $T_{\mathrm{F}}(\protect%
\omega_{\mathrm{p}}=\protect\omega_{\mathrm{C}})$ as a function of $(\Delta_{%
\mathrm{AC}},g_0)$, with $(\protect\kappa_{\mathrm{i}},h)$ determined from
fits to experimental data. }
\label{figB}
\end{figure}

Explicitly, under conditions of critical coupling, $T_{\mathrm{F}}$ can be
expressed as
\begin{eqnarray}
T_{\mathrm{F}} \big|_{\omega_{\mathrm{p}}=\omega_{\mathrm{C}}}= \frac{%
4\kappa_{\mathrm{i}}^2\left|g_{\mathrm{tw}}\right|^4+h^2\left( g_{\mathrm{tw}%
}^2+(g_{\mathrm{tw}}^\ast)^2\right)^2} {\left[\gamma \left(h^2+\kappa^2%
\right)+2\kappa\left|g_{\mathrm{tw}}\right|^2\right]^2 + \left[\Delta_{%
\mathrm{AC}}\left(h^2+\kappa^2\right)-h\left( g_{\mathrm{tw}}^2+(g_{\mathrm{%
tw}}^\ast)^2\right)\right]^2} \, .  \label{Tf}
\end{eqnarray}
This is simply a Lorentzian centered at
\begin{equation}
\Delta_{\mathrm{AC}}^{center} = \frac{h}{h^2+\kappa^2} \left( g_{\mathrm{tw}%
}^2+(g_{\mathrm{tw}}^\ast)^2\right) ,
\end{equation}
(note that when averaged over the azimuthal coordinate $x$, $\Delta_{\mathrm{%
AC}}^{center}$ approaches zero), and with half-width
\begin{equation}
\beta = \gamma + \frac{2\kappa\left|g_{\mathrm{tw}%
}\right|^2+h\left( g_{\mathrm{tw}}^2+(g_{\mathrm{tw}}^\ast)^2\right)} {%
h^2+\kappa^2} \simeq \frac{2\left|g_{\mathrm{tw}}\right|^2}{\kappa} = \frac{%
|g_0|^2}{\kappa} \, ,
\end{equation}

\noindent assuming $2|g_{\mathrm{tw}}|^2/\kappa \gg \gamma $ and $h/\kappa
\ll 1$. The above result for $T_{\mathrm{F}}$, averaged over the azimuthal
coordinate $x$, is shown in Fig.~\ref{figB} as a function of $\Delta_{%
\mathrm{AC}}$ and $g_{\mathrm{0}}=\sqrt{2}\, g_{\mathrm{tw}}$. Note that the
half-widths of the curves $T_{\mathrm{F}}(\Delta_{\mathrm{AC}})$ are
well-approximated by $g_{\mathrm{0}}^2/\kappa$. Radial and temporal
averaging leads to substantial narrowing of the curves $T_{\mathrm{F}%
}(\Delta_{\mathrm{AC}})$, producing a $\beta_{\mathrm{effective}}$ which,
while significantly smaller than the $\beta$ defined in Eq. (5), maintains
the dependence on the maximum coupling strength $g_{\mathrm{0}}^m$, as
illustrated in Fig.~4.
Therefore, given knowledge of $\kappa $ and $\gamma$, a measurement of the
dependence of $T_{\mathrm{F}}$ on $\Delta_{\mathrm{AC}}$ yields $g_{\mathrm{0%
}}$ directly.

Note that the linear model presented here is adequate for the regime of our
current experiment, as we have confirmed by numerical solutions of the full
master equation. For the probe resonant with the cavity frequency, $\omega _{%
\mathrm{p}}=\omega _{\mathrm{C}}$, the population in the atomic excited
state remains negligible for the conditions of our experiment.\newline

It is interesting to contrast our model and the conventional Jaynes-Cummings
model, which considers a two-level system coupled to only one
electromagnetic field mode (for a given polarization), as is typically the
case with Fabry-Perot resonators. As described above, unlike the two dressed
states of the Jaynes-Cummings model, there are three eigenstates in our
case. In the specific cases of $kx=0,\pi/2$, two dressed states correspond
to the vacuum-Rabi sidebands resulting from the coupling of the atom to one
normal mode, while the third is essentially the normal mode that remains
uncoupled with the atom. In principle, given that the coupling to the atom
is stronger than the intermode coupling $h$, the definition of the normal
modes could be chosen so the atom is at the node of one, and at the
anti-node of the other for every value of $kx$. Thus, unlike the situation
with Fabry-Perot resonators, where the atom could experience strong coupling
at the anti-nodes of the standing wave and remain uncoupled to the cavity at
the nodes of the standing wave, here the atom always obtains roughly the
same degree of coupling with the cavity (either through one mode, the other
or both). Accordingly, roughly half of the light in the cavity always
remains uncoupled to the atom, leading to the third eigenvalue around zero
detuning, as described in Fig~\ref{fig5}. Note also that the presence of the
atom redistributes the intensity between the two normal modes \cite%
{domokos00}. Finally, the existence of two normal modes leads to the factor
of $\sqrt{2}$ between the coupling constant in our case and the one
predicted by the Jaynes-Cummings model for a traveling wave $g_{\mathrm{0}}=%
\sqrt{2}g_{\mathrm{0}}^{\mathrm{tw}}$. Note that for the single-mode
Janes-Cummings model, Eq.~(\ref{Tf}) takes the form
\begin{eqnarray}
T_F = \frac{\left( \left|g_{\mathrm{tw}}\right|^2/\kappa \right)^2} {%
\left(\gamma +\left|g_{\mathrm{tw}}\right|^2/\kappa \right)^2+\Delta_{%
\mathrm{AC}}^2} \, ,
\end{eqnarray}
illustrating the same Lorentzian dependence on $\Delta_{\mathrm{AC}}$, with
half-width $\beta_{\mathrm{JC}} \sim \left|g_{\mathrm{tw}}\right|^2/\kappa$
for $\left|g_{\mathrm{tw}}\right|^2/\kappa\gg\gamma $.

\section{Calculation of the coherent coupling parameter $g_{\mathrm{0}}$}

For our particular toroidal resonator with major diameter $D\simeq 44$ $\mu $%
m and minor diameter $d\simeq 6$ $\mu $m, we find numerically the lowest
order traveling-wave mode functions $\,\phi _{\mathrm{tw}}^{\pm }(\rho ,x,z)$
of the resonator \cite{spillane05}, from which follows the coupling\
parameters $g_{\mathrm{0}}^{\mathrm{tw}}$ and $g_{0}$. For the $%
6S_{1/2},F=4,m_{F}=4\longrightarrow 6P_{3/2},F^{\prime }=5^{\prime
},m_{F}^{\prime }=5^{\prime }$ transition of the $D_{2}$ line of atomic
Cesium, we find $g_{\mathrm{0}}^{\mathrm{tw}}/2\pi =80$ MHz and thus $g_{%
\mathrm{0}}/2\pi =\sqrt{2}\times 80$ MHz. However, a circularly polarized
field is required for coupling to this transition while the toroidal
resonator supports linear polarization. Hence, for atoms uniformly
distributed over the set of Zeeman states $\{m_{F}\}$ in the $F=4$ ground
state, we calculate $g_{\mathrm{0}}$ from an average over Clebsch-Gordon
coefficients for $\Delta m_{F}=0$ transitions for $6S_{1/2},F=4%
\longleftrightarrow 6P_{3/2},F^{\prime }=5^{\prime }$, leading to $g_{%
\mathrm{0}}/2\pi =70$ MHz, which is the value utilized in Fig. 5 above and
quoted in the main text.\newline

\newpage

\textbf{Current addresses --}

$^{a}$ TA -- Department of Applied Physics, The University of Tokyo, Tokyo,
Japan

$^{b}$WPB -- Physics Department, University of Otago, Dunedin, New Zealand

$^{c}$ ASP -- Department of Physics, University of Auckland, Auckland, New
Zealand

$^{d}$ TJK -- Max Planck Institute of Quantum Optics, Garching, Germany

\newpage

\end{document}